\documentclass[prd,nofootinbib,amsfonts,notitlepage]{revtex4-1}
\usepackage[utf8]{inputenc}
\usepackage[T1]{fontenc}
\usepackage{hyperref}
\usepackage{graphicx,bm,amsmath,color}
\DeclareMathOperator{\arccosh}{arccosh}

\usepackage{bbold}
\usepackage{wasysym}
\usepackage{verbatim}
\usepackage{amssymb}
\usepackage{epstopdf}
\usepackage{animate}
\usepackage{xmpmulti}
\usepackage{centernot}
\usepackage{multirow}
\usepackage{tikz}
\usepackage{graphicx}
\usepackage{float}
\usepackage{mathtools}
\usepackage{comment}
\usepackage{pgfplots}
\pgfplotsset{width=10cm,compat=1.9}
\usepgfplotslibrary{external}
\makeatletter

\newcommand{\be}{\begin{equation}}
\newcommand{\ee}{\end{equation}}
\newcommand{\beq}{\begin{eqnarray}}
\newcommand{\eeq}{\end{eqnarray}}
\newcommand{\ba}{\begin{align}}
\newcommand{\ea}{\end{align}}

\begin{document}

\title{Constraints on the deformation scale of a geometry in the cotangent bundle}
\author{J.J. Relancio}
\affiliation{Departamento de F\'{\i}sica Te\'orica and Centro de Astropartículas y Física de Altas Energías (CAPA),
Universidad de Zaragoza, Zaragoza 50009, Spain}
\email{relancio@unizar.es}
\author{S. Liberati}
\affiliation{SISSA, Via Bonomea 265, 34136 Trieste, Italy and INFN, Sezione di Trieste;\\ IFPU - Institute for Fundamental Physics of the Universe, Via Beirut 2, 34014 Trieste, Italy}
\email{liberati@sissa.it}

\begin{abstract}
There are several studies proposing phenomenological consequences of a deformation of special and general relativity. Here, we cast novel constraints on the deformation parameter of a metric in the cotangent bundle accounting for a curved momentum space. In an expanding universe, we study three possible observations that could restrict our model, focusing on the deformations of  velocity, redshift and luminosity distance, which in the aforementioned framework, depend on the energy of the particles. We find that for an energy dependent velocity there would be no time delay for massless particles since also the observed distance to the source depends on the energy. For the redshift and luminosity distance we see that a scale of the order of some keV could be compatible with our model. This shows that the constraints on the high-energy scale parametrizing the momentum dependent deviation from a Friedmann-Robertson-Walker metric are at the moment weak due to the fact that the precision (rather than energies) needed in the observational constraints are extremely high. However, this is not the case when considering the synchrotron radiation. Indeed, the observation of such emission from the Crab Nebula, for deformations leading to subluminal propagation at high energies, leads to a constraint for the high-energy scale of the order of $1$ PeV.
\end{abstract}

\maketitle

\section{Introduction}
The search for a quantum gravity theory (QGT) which would reconcile quantum field theory (QFT) and general relativity (GR) has been the focus of intense investigations for many years by now. In the literature, there are several theoretical attempts for achieving such goal, like string theory~\cite{Mukhi:2011zz,Aharony:1999ks,Dienes:1996du}, loop quantum gravity~\cite{Sahlmann:2010zf,Dupuis:2012yw}, supergravity~\cite{VanNieuwenhuizen:1981ae,Taylor:1983su}, or causal set \mbox{theory~\cite{Wallden:2013kka,Wallden:2010sh,Henson:2006kf}}. One of the most noticeably features that these theories have in common is the presence of a minimum length~\cite{Gross:1987ar,Amati:1988tn,Garay1995}, normally associated with the Planck length $\ell_P \sim 1.6\times 10^{-33}$\,cm, which implies a characteristic energy scale $M_P \sim 1.22\times 10^{16}$\,TeV. 

One can distinguish different scenarios depending on the Lorentz symmetry fate. For example, Lorentz invariance is (at least in principle) maintained in loop quantum gravity~\cite{Gambini:1998it}, in which a foam  arises~\cite{Wheeler:1955zz,Rovelli:2002vp,Ng:2011rn,Perez:2012wv} (that can be interpreted as a ``quantum'' spacetime), and in causal set theory and string theory, in which non-locality effects appear~\cite{Belenchia:2014fda,Belenchia:2015ake}. The main problem is that there is a lack of phenomenology in these theories, making it very difficult to determine which of them is the correct approach to a QGT. However, some phenomenological studies have been carried out in particle accelerators~\cite{Anchordoqui:2008eu,Nayak:2009ke,Calmet:2010nt,Biswas:2014yia} (considering that the characteristic scale could be of the order of some TeV) and also on astrophysical sources~\cite{Mattingly:2005re,Liberati:2013xla,Visinelli:2017bny,Pardo:2018ipy,Calcagni:2019ngc} (time delay of massless particles, luminosity distance...). 

Nevertheless, there are frameworks in which Lorentz symmetry is only a low energy feature broken in the ultra-violet range of energies. In these scenarios of Lorentz invariance violation  (LIV)~\cite{Colladay:1998fq}, there are many studies constraining the scale of Lorentz breaking (see~\cite{Kostelecky:2008ts,Liberati:2013xla} and references therein) in particle dispersion relations.

However, a hard violation of Lorentz invariance is not the only possibility to go beyond special relativity (SR). In doubly special relativity (DSR)~\cite{AmelinoCamelia:2008qg}, the symmetries are deformed for high energies, not broken. This means that a relativity principle holds, making  all inertial observers describe the same physics. While in LIV scenarios there are many testable effects due to the modification of the kinematics~\cite{Liberati:2013xla} (modification in thresholds or allowed processes that are forbidden in SR~\cite{Jacobson:2002hd}), this is not the case in DSR. This is due to the fact that LIV corrections are of order $E^n/\Lambda^{n-2}$ with $n \geq 3$, whilst in DSR the modification of the thresholds are generally very small and the processes that are forbidden in SR are not allowed in this scheme, due to the presence of a relativity principle. Then, since the scale of deformation is considered to be of the order of the Planck scale, the usual phenomenological window for testing DSR is a time delay of photons~\cite{Amelino-Camelia1998,AmelinoCamelia:2011cv,Loret:2014uia,Carmona:2017oit,Carmona:2018xwm,Carmona:2019oph}.

Moreover, there are several studies in both LIV~\cite{Kostelecky:2011qz,Barcelo:2001cp,Weinfurtner:2006wt} and DSR~\cite{Liberati:2004ju,Aloisio:2005qt,Aloisio:2006nd,Girelli:2006fw,Amelino-Camelia:2014rga} considering a modification of the Riemannian geometry known as Finsler geometry (this is a particular case of Lagrange space geometries~\cite{2012arXiv1203.4101M}). In this frameworks the metric, constructed from a deformed dispersion relation, can depend on the tangent vectors besides the space-time coordinates. In addition, DSR kinematics have been understood from the geometry of a curved momentum space~\cite{AmelinoCamelia:2011bm,Lobo:2016blj}. Indeed, in~\cite{Carmona:2019fwf} it was rigorously shown that a de Sitter momentum space leads to $\kappa$-Poincaré kinematics identifying the isometries and the squared distance of the metric with the main ingredients of the kinematics: translations, Lorentz isometries and the square of the distance can be interpreted as the deformed composition law, deformed Lorentz transformations and deformed dispersion relation respectively (the last two facts where previously contemplated in Refs.~\cite{AmelinoCamelia:2011bm,Lobo:2016blj}).

Nevertheless, in order to study a possible time delay for massless particles, one needs to take into account the expansion of the universe. Then, it is mandatory to have a theory combining DSR with a curved spacetime. In the existent literature, there are several works trying to gather a relativistic deformed kinematics with a curved spacetime~\cite{Cianfrani:2014fia,Rosati:2015pga}. In~\cite{Relancio:2020zok}, we proposed a simple generalization of~\cite{Carmona:2019fwf} in order to consider a curved  spacetime, so obtaining a metric depending on space-time and momentum variables. This approach differs from other works in the literature~\cite{Barcaroli:2015xda,Barcaroli:2016yrl,Letizia:2016lew,Barcaroli:2017gvg} in the framework of Hamiltonian geometry (the Hamiltonian version of the Finsler space in which the metric can depend on the cotangent vectors~\cite{2012arXiv1203.4101M})  in that the starting point is a metric in momentum space, not a deformed dispersion relation. Indeed in the latter case there is an ambiguity given that considering the so-called ``classical basis'' of $\kappa$-Poincaré~\cite{Borowiec2010}, in which the dispersion relation is the usual one of SR, no momentum dependence of the metric appears to arise.

In Ref.~\cite{Relancio:2020zok} some phenomenological aspects were discussed for a Friedmann-Robertson-Walker universe. In particular, we found that there is a dependence on the energy for the velocity, redshift and luminosity distance. In this work, we use experimental data in order to obtain constraints on the high-energy scale characterizing the momentum dependence of the metric.

Moreover, possible phenomenological consequences in DSR  regarding the synchrotron radiation were studied in~\cite{AmelinoCamelia:2010pd,AmelinoCamelia:2011bq}. While in LIV scenarios a critical frequency is obtained when the velocity of electrons is always subluminal~\cite{Jacobson:2005bg}, in~\cite{AmelinoCamelia:2010pd} it was argued that a deeper study is required in order to check if the same effect takes place in DSR. We will show that from our metric formalism a similar effect appears in this case, leading to a completely different (and weaker) constrain on the high-energy scale parametrizing the momentum deviation from the Minkowski metric. 

The structure of the paper is as follows. In Sec.~\ref{sec:cotangent} we summarize how a metric in the cotangent bundle containing  a curved momentum space and a nontrivial spacetime can be constructed, and the phenomenological results obtained for the particular case of Friedmann-Robertson-Walker universe. After that, in Sec.~\ref{sec:expanding_universe} we use recent experimental data in order to constrain the scale of deformation. We find that, despite having a velocity depending on momentum pointing to a possible time delay for photons with different energies, the distance to the source depends also on the energy, leading to an absence of time delay. However, for the redshift and luminosity distance cases, we find that a scale of some keV could be compatible with recent experiments. In Sec.~\ref{sec:synchrotron} we see that a maximum frequency appears for synchrotron radiation, leading to a constrain on the deformation scale of the order of 1 PeV. Finally, we end with conclusions and future prospects in Sec.~\ref{sec:conclusions}.

\section{Cotangent geometry in a nutshell}
\label{sec:cotangent}

In this section we review the main ingredients of our proposal for a metric in the cotangent bundle, gathering a de Sitter momentum space and any curved spacetime. We start by describing the free propagation of particles in SR from an action and see how this action can be generalized in order to depict the propagation in a curved spacetime. Let us then start with the action for a particle in SR
\begin{equation}
S\,=\,\int{d\tau \left(\dot{x}^\mu k_\mu-\mathcal{N} \left(C(k)-m^2\right)\right)}\,,
\label{eq:SR_action}
\end{equation}
where $C(k)=k_\alpha \eta^{\alpha\beta }k_\beta$ is the SR dispersion relation and the dot represents the derivative with respect to the proper time of the particle, $\tau$. Let us now note that one can obtain the geodesic motion in a curved geometry by just rewriting Eq.\eqref{eq:SR_action} as
\begin{equation}
S\,=\,\int{d\tau \left(\dot{x}^\mu k_\mu-\mathcal{N} \left(C(\bar{k})-m^2\right)\right)}\,,
\label{eq:GR_action}
\end{equation}
where $\bar{k}_\alpha=\bar{e}^\nu_\alpha (x) k_\nu$, with $\bar{e}^\nu_\alpha(x)$ defined as the inverse of the tetrad of the space-time metric  $e_\nu^\alpha(x)$, satisfying
\begin{equation}
g^x_{\mu\nu}(x)\,= \, e^\alpha_\mu (x) \eta_{\alpha\beta} e^\beta_\nu (x)\,,
\label{eq:metric-st}
\end{equation}
while the dispersion relation is given by  
\begin{equation}
C(\bar{k})\,=\,\bar{k}_\alpha \eta^{\alpha\beta }\bar{k}_\beta\,=\,k_\mu g_x^{\mu\nu}(x) k_\nu\,.
\label{eq:cass_GR}
\end{equation}
It is easy to check that the worldlines obtained through this action are the same that one would obtain in a metric theory of gravity via the geodesic equation.

Let us now define the metric in momentum space (with no dependence on the space-time coordinates~\cite{Carmona:2019fwf})
\begin{equation}
g_k^{\mu\nu}(k)\,=\, \bar{\varphi}_\alpha^\mu(k)\eta^{\alpha\beta}\bar{\varphi}_\beta^\nu(k) \,,
\end{equation}
where $\bar{\varphi}^\mu_\alpha(k)$ is the tetrad in momentum space, so that the line element in momentum space is 
\begin{equation}
d\sigma^2\,=\,dk_{ \alpha}g_k^{\alpha\beta}(k)dk_{ \beta}\,=\,dk_{ \alpha}\bar{\varphi}^\alpha_\gamma(k)\eta^{\gamma\delta}\bar{\varphi}^\beta_\delta(k)dk_{ \beta}\,.
\label{eq:line_m1}
\end{equation}
Since in the step of going from SR to a curved spacetime one replaces in the action $k\to \bar{k}$, we propose that the line element in momentum space, when a curvature in spacetime is considered, becomes
\begin{equation}
d\sigma^2\,\coloneqq\,d \bar{k}_{\alpha}g_{\bar{k}}^{\alpha\beta}( \bar{k})d \bar{k}_{ \beta}\,=\,dk_{\mu}g^{\mu\nu}(x,k)dk_{\nu}\,,
\label{eq:momentum_line_element}
\end{equation}
being 
\begin{equation}
g_{\mu\nu}(x,k)\,=\,\Phi^\alpha_\mu(x,k) \eta_{\alpha\beta}\Phi^\beta_\nu(x,k)\,,
\label{eq:cotangent_metric_tetrads}
\end{equation}
where 
\begin{equation}
\Phi^\alpha_\mu(x,k)\,=\,e^\lambda_\mu(x)\varphi^\alpha_\lambda(\bar{k})\,.
\label{eq:tetrad_cotangent}
\end{equation}
and where $\varphi^\alpha_\lambda(k)$ is the inverse of the tetrad in momentum space.

If the base manifold has dimension $n$, the cotangent bundle manifold has dimension $2n$. On the cotangent bundle, one can consider its tangent bundle, which has also $2n$ dimensions. There is in the last construction a remarkable distribution generated by $\partial/\partial k_\mu$, which is called the vertical distribution $V$, with dimension $n$. As it is shown in~\cite{2012arXiv1203.4101M} one can define a  nonlinear connection $N$ (also called horizontal distribution), supplementary to the vertical distribution, and also with dimension $n$.

One can construct an adapted basis for the horizontal distribution
\begin{equation}
\frac{\delta}{\delta x^\mu}\, \doteq \,\frac{\partial}{\partial x^\mu}+N_{\nu\mu}(x,k)\frac{ \partial}{\partial k_\nu}\,,
\label{eq:delta_derivative}
\end{equation}  
where $N_{\mu \nu}$ are called the coefficients of the nonlinear connection. The choice of these coefficients is not unique but, as it is shown in~\cite{2012arXiv1203.4101M}, there is one and only one choice of nonlinear connection coefficients that leads to a space-time affine connection which is metric compatible (in the sense of covariant derivative, as we show below) and torsion free.
In GR, the coefficients of the nonlinear connection are given by
\begin{equation}
N_{\mu\nu}(x,k)\, = \, k_\rho \Gamma^\rho_{\mu\nu}(x)\,,
\label{eq:nonlinear_connection}
\end{equation} 
being $\Gamma^\rho_{\mu\nu}(x)$ the affine connection. 

The line element in momentum space Eq.~\eqref{eq:momentum_line_element} is just a part of the line element in the whole phase space (see Ch.~4 of  Ref.~\cite{2012arXiv1203.4101M}), in which one can define a line element for the cotangent bundle
\begin{equation}
\mathcal{G}\,=\, g_{\mu\nu}(x,k) dx^\mu dx^\nu+g^{\mu\nu}(x,k) \delta k_\mu \delta k_\nu\,,
\label{eq:line_element_ps} 
\end{equation}
where 
\begin{equation}
\delta k_\mu \,=\, d k_\mu - N_{\nu\mu}(x,k)\,dx^\nu\,. 
\end{equation}

From this line element we can define two different types of curves. The vertical curves, which allow us to move along a fiber for a fixed space-time point, lead to the line momentum element we have written previously
\begin{equation}
\mathcal{E}\,=\,d\sigma^2\,=\,g^{\mu\nu}(x,k) d k_\mu d k_\nu\,.
\label{eq:line_element_p} 
\end{equation}
As we have mentioned, from this we can define the Casimir function to be the squared distance in momentum space for a fixed space-time point $x_0$: the distance from $(x_0,0)$ to $(x_0,k)$. In GR, in which the metric does not depend on the momenta, this leads to Eq.~\eqref{eq:cass_GR}.     

The horizontal curves lead to the usual geodesics in GR, indeed the line element 
\begin{equation}
\mathcal{E}\,=\,ds^2\,=\, g_{\mu\nu}(x,k) dx^\mu dx^\nu\,,
\label{eq:line_element_s} 
\end{equation}
implies a geodesic equation of the form 
\begin{equation}
\frac{\delta k_\lambda}{\delta \tau}\,=\,\frac{dk_\lambda}{d\tau}-N_{\sigma\lambda} (x,k)\frac{dx^\sigma}{d\tau}\,=\,0\,.
\end{equation}
The coefficients of the nonlinear connection are then giving the evolution of the momenta as a function of $\tau$. Moreover, from the previous line element one can obtain the geodesic equation~\cite{2012arXiv1203.4101M}
\begin{equation}
\frac{d^2x^\mu}{d\tau^2}+H^\mu_{\nu\sigma}(x,k)\frac{dx^\nu}{d\tau}\frac{dx^\sigma}{d\tau}\,=\,0\,,
\label{eq:horizontal_geodesics}
\end{equation} 
where 
\begin{equation}
H^\rho_{\mu\nu}(x,k)\,=\,\frac{1}{2}g^{\rho\sigma}(x,k)\left(\frac{\delta g_{\sigma\nu}(x,k)}{\delta x^\mu} +\frac{\delta g_{\sigma\mu}(x,k)}{\delta  x^\nu} -\frac{\delta g_{\mu\nu}(x,k)}{\delta x^\sigma} \right)\,,
\label{eq:affine_connection_st}
\end{equation} 
is the affine connection of spacetime. When the metric does not depend on the momenta one obtains the same result of GR.

In~\cite{Relancio:2020zok} we have considered the following choice of momentum metric
\begin{equation}
g^k_{00}(k)\,=\,1\,,\qquad g^k_{0i}(k)\,=\,0\,, \qquad  g^k_{ij}(k)\,=\,\eta_{ij}\, e^{\mp 2k_0/\Lambda}\,,
\label{eq:dS_metric3}
\end{equation}
where $\Lambda$ plays the role of the parameter of the deformation. Both signs describe a de Sitter momentum space, where we always regard $\Lambda$  as a positive energy scale~\footnote{As it was shown in~\cite{Weinberg:1972kfs}, this particular form can not be obtained for the anti de Sitter case just interchanging the sign before $\Lambda$. Since the scalar of curvature is proportional to the inverse of $\Lambda$ squared, in these particular coordinates both signs represent a de Sitter space.}.

From this momentum metric, we obtained three important phenomenological results~\cite{Relancio:2020zok}:
\begin{enumerate}
	\item For a Friedmann-Robertson-Walker universe, the metric in the cotangent bundle is
	\begin{equation}
g_{00}(x,k)\,=\,1\,,\qquad g_{0i}(x,k)\,=\,0\,, \qquad  g_{ij}(x,k)\,=\,\eta_{ij}\,R^2(x^0)\,  e^{\mp 2k_0/\Lambda}\,,
\label{eq:FRW_metric}
\end{equation}
where $R$ is the usual scale factor, and the velocity in these coordinates is (in $1+1$ dimensions)~\cite{Relancio:2020zok}
\begin{equation}
\frac{d x^1}{dx^0}\,=\, \frac{e^{\pm k_0/\Lambda}}{R(x^0)}\,.
\label{eq:velocity}
\end{equation}
	\item In this metric, there is a modification on the redshift, so that higher-energy particles suffer a larger redshift~\cite{Relancio:2020zok}
	\begin{equation}
1+z(E)\,=\,(1+z(0))\left( 1 \pm \frac{E}{\Lambda }\left(\frac{1}{R(t_0)}-\frac{1}{R(t_1)}\right)\right)\,=\,(1+z(0))\left( 1 \pm \frac{E}{\Lambda }z(0)\right)\,,
	\label{eq:redshift}
\end{equation}
where in the last step we have used that 
\begin{equation}
1+z(0)\,=\, \frac{R(t_0)}{R(t_1)}\,=\,\frac{1}{R(t_1)}\,\implies \frac{1}{R(t_0)}-\frac{1}{R(t_1)}\,=\, z(0)\,,
\label{eq:redshift_low}
\end{equation}
	since we are considering that $R(t_0)=1$.
	\item Also, we found that the luminosity distance becomes energy dependent in a similar way~\cite{Relancio:2020zok}
\begin{equation}
\frac{d_L (0)}{d_L (E)}\,=\,\left(\frac{1+z(0)}{1+z(E)}\right)^2\,=\, 1\mp \frac{2E}{\Lambda }z(0)\,.
	\label{eq:luminosity_distance}
\end{equation}
\end{enumerate}

\section{Cosmological constraints}
\label{sec:expanding_universe}
Let us now investigate the phenomenological implications of our framework in order to constrain it on the basis of the available observations.

\subsection{Time delays}
\label{sec:time_delays}

Since we have a momentum dependent metric, particles with different energies probe different spacetimes. However, as anticipated, this does not lead in our framework to the usual time delays between photons of different energy as normally predicted in DSR~\cite{Amelino-Camelia1998,AmelinoCamelia:2011cv,Loret:2014uia} and Lorentz breaking scenarios~\cite{Vasileiou:2015wja,Ellis:2018lca,Abdalla:2019krx,Xu:2018ien}. We start by considering the particular DSR scenario in which there is a curved momentum space but a flat spacetime. Then, for a metric depending only on momentum coordinates as in Eq.~\eqref{eq:dS_metric3}, the line element in spacetime is (in $1+1$ dimensions)\footnote{We use the minus sign through Eq.\eqref{eq:dS_metric3} here, but the same results can be recover in a straight-forward fashion for the opposite sign.}
\begin{equation}
ds^2\,=\, dt^2-dx^2 e^{-2 k_0/\Lambda}\,.
\label{eq:line_flat}
\end{equation}
Let us suppose a source at a distance $r$ from our laboratory emitting light at different frequencies. For a low energy photon we can neglect the factor $e^{-2 k_0/\Lambda}$ (given that $k_0/\Lambda\approx 0$) so that $r=x_l$, where the $x_l$ is the space coordinate of a low energy photon. However, this is not the case for a high energy photon.  From the line element then
\begin{equation}
ds^2\,=\, dt^2-dr^2\,=\,dt^2-dx_l^2\,=\, dt^2-dx_h^2 \,e^{-2 k_0/\Lambda}\,,
\end{equation}
which is tantamount to say that
\begin{equation}
r\,=\, x_l\,=\,x_h\, e^{-k_0/\Lambda}\,,
\label{eq:dS_proper_distance}
\end{equation}
being $r$ the real distance between the source and us.  Then, if the distance is given by Eq.~\eqref{eq:dS_proper_distance}, the velocity in these coordinates is 
 \begin{equation}
\frac{dr}{dt}\,=\,\frac{dx_h}{dt}\,e^{-k_0/\Lambda}\,=\,1\,,
\end{equation}
where we have used the constancy of the energy due to the independence of the metric on the space-time coordinates and that from the metric element for photons one finds
\begin{equation}
ds^2\,=\,0\,\implies 1\,=\,\frac{dx_h^2}{dt^2} e^{-2 k_0/\Lambda}\,.
\end{equation}
Then, we have found that in flat spacetime the speed of photons is independent on their energy, which agrees with others results in the literature~\cite{Carmona:2018xwm,Carmona:2019oph}\footnote{In~\cite{Carmona:2018xwm,Carmona:2019oph} it was followed a different approach to the usual DSR studies, both for the  noncommutativity of spacetime~\cite{Amelino-Camelia1998,AmelinoCamelia:2011cv,Loret:2014uia} and for the geometrical approaches of Finsler~\cite{Liberati:2004ju,Aloisio:2005qt,Aloisio:2006nd,Girelli:2006fw,Amelino-Camelia:2014rga} and Hamilton~\cite{Barcaroli:2015xda,Barcaroli:2016yrl,Letizia:2016lew,Barcaroli:2017gvg} spaces.}.

Now we can generalize the previous study to the curved space-time case in a direct way. The line element for a momentum modification of the Friedmann-Robertson-Walker metric is given by the metric~\eqref{eq:FRW_metric}
\begin{equation}
ds^2\,=\, dt^2-dr^2 R^2(t)\,=\,dt^2- dx_l^2\,R^2(t)\,=\, dt^2-dx_h^2\, R^2(t)  \,e^{-2 k_0(t)/\Lambda}\,,
\end{equation}
where the zero component of the momentum now depends on the time due to the redshift. Therefore, one can find for photons 
\begin{equation}
dx_h\,=\, \frac{e^{k_0(t)/\Lambda}}{R(t)} dt\,.
\end{equation}
Then, if we define the (differential) real distance to the source as 
\begin{equation}
dr\,=\, dx_l \,=\,dx_h\, e^{-k_0(t)/\Lambda}\,,
\label{eq:dS-FRW_proper_distance}
\end{equation}
we find that the velocity defined as
 \begin{equation}
\frac{dr}{dt}\,=\,\frac{dx_h}{dt}\, e^{-k_0(t)/\Lambda}\,=\,\frac{1}{R(t)}\,,
\end{equation}
is independent of the photon energy. With this we show that there is no time delay for massless particles with our proposal, which is in agreement with the actual experimental data~\cite{Vasileiou:2015wja,Ellis:2018lca,Abdalla:2019krx}.

\subsection{Redshift}
\label{sec:redshift}

In this subsection, we use the redshift of several active galactic nuclei (we use the same source numeration of Ref.~\cite{Rangel_2012}) obtained from the visible and x-ray spectra~\cite{Simmonds_2018}, that can be found in Table~\ref{table:1}. Since the energy of x-rays are between $2$ and $7$ keV and the visible spectrum energy are of the order of a few eV~\cite{Pello:1996fk}, we can use Eq.~\eqref{eq:redshift} in order to establish the lower bound on the scale $\Lambda$: 
	\begin{equation}
1+z(E)\,=\,(1+z(0))\left( 1 \pm \frac{E}{\Lambda }z(0)\right)\implies \frac{1+z(E)}{1+z(0)}-1\,=\, \pm \frac{E}{\Lambda} z(0) \,.
\end{equation}
From here, it is easy to find
 \begin{equation}
\Lambda\,=\,E \frac{ z(0)(1+z(0))}{\sqrt{(z(E)-z(0))^2}} \,.
\end{equation}
Now we can cast a constraint for $\Lambda$ by requiring 
 \begin{equation}
\Lambda\,>\,E_\text{obs} \frac{ z_\text{obs}(0)(1+z_\text{obs}(0))}{|z_\text{obs}(E)-z_\text{obs}(0)|} \,,
\end{equation}
obtaining the minimum value of $\Lambda$ compatible with all the data with the errors. For each source we obtain the lower value of $\Lambda$ compatible with the experimental results. Then, averaging the scale obtained for all cases, we find that $\Lambda>49$ keV  (with $E_\text{obs}=7$ keV) would be compatible with present observational data. 

\begin{table}[H]
\centering
\begin{tabular}{||c c c||} 
 \hline
 Source & X-ray redshift & Visible redshift \\ [0.5ex] 
 \hline\hline
 4 & $0.31^{+0.02}_{-0.03}$ & $0.28^{+0.01}_{-0.01}$  \\ 
 \hline
 5 & $1.89^{+1.20}_{-1.46}$ & $2.88^{+1.98}_{-1.17}$\\
 \hline
 6 & $0.78^{+0.48}_{-0.48}$ & $1.06^{+0.10}_{-0.02}$ \\
 \hline
 7 & $3.35^{+0.05}_{-0.05}$ & $2.97^{+0.08}_{-0.11}$  \\
 \hline
 13 & $1.97^{+0.93}_{-1.36}$ & $2.60^{+0.16}_{-0.19}$  \\ 
\hline
 16 & $1.81^{+0.16}_{-1.03}$ & $1.85^{+0.09}_{-0.09}$  \\
\hline
\end{tabular} 
\caption{Redshift obtained from x-ray and visible light for different sources}
\label{table:1}
\end{table}

\subsection{Luminosity distance}
\label{sec:luminosity_distance}
In this subsection, we use the combined results of the gravitational waves and gamma rays from a binary neutron star merger,   GW170817~\cite{TheLIGOScientific:2017qsa} and GRB170817A~\cite{Goldstein:2017mmi}, respectively. From the first, the luminosity distance of the gravitational wave is $40^{+8}_{-14}$ Mpc, while for the gamma ray burst $42.9 \pm 3.2$ Mpc. There is also a difference in the energy of both signals: the energy of the gravitational wave is practically negligible ($10^{-15}$ keV) while the energy of the photon peak is around $300$ keV (we are considering the data of~\cite{Monitor:2017mdv} and references therein). 
 The redshift $z=0.00968$ is computed with photons with energy of some eV~\cite{Jones_2009} so we can use Eq.~\eqref{eq:luminosity_distance}:
\begin{equation}
\frac{d_L (E)-d_L (0)}{d_L (0)}\,=\,\pm 2\frac{E}{\Lambda} z(0)\,,
	\label{eq:luminosity_distance_number}
\end{equation}
and then
\begin{equation}
\Lambda\,>\,\frac{ 2  \,E_\text{obs} \, z_\text{obs}(0)\, d_{L\,\text{obs}} (0)}{\sqrt{d_{L\,\text{obs}}^2 (E)-d^2_{L\,\text{obs}} (0)}} \,,
	\label{eq:luminosity_distance_number_2}
\end{equation}
finding that $\Lambda>8$ keV would be compatible with the data.

In summary, we see that there would be no time delay in our proposal since the distance depends also on the energy, in such a way that there is a cancellation of effects (distance and velocity dependence on the energy). Also, when using cosmological data in order to constrain the high-energy scale, we find that a momentum dependent redshift and luminosity distance lead to very weak bounds on the scale, being these of the order of some tens of keV.  

This shows that from  cosmological phenomenology we are not able to cast strong constraints on the high-energy scale deforming the usual Friedmann-Robertson-Walker metric. 

\section{Astrophysical constraints: synchrotron radiation}
\label{sec:synchrotron}

In this section we will study the behavior of the synchrotron frequency with a metric formalism corresponding to a de Sitter momentum space with no dependence on the space-time coordinates. We will compute the  critical frequency of the synchrotron radiation with the metric~\eqref{eq:dS_metric3}. As we will see, when the velocity of electrons are always subluminal, we find a maximum frequency of the emitted photons. 

In order to obtain the critical frequency of the synchrotron radiation, we use the same procedure followed in Ref.~\cite{Jacobson:2005bg} in the LIV context. An electron of energy $E$ emits radiation in a cone of some opening angle $\delta (E)$. The cone sweeps past a distant observer as the electron moves on a circle of radius $R(E)$ through the emission angle. The electron speed is $v(E)$, so the time it takes to orbit through the angle $\delta (E)$ is $\Delta t=R(E)\delta (E)/v(E)$. The light from the leading edge of the cone travels a distance $c(\omega) \Delta t$ while the electron travels the distance $v(E) \Delta t$ along the circular trajectory pointing towards the observer. Hence the spatial width of the pulse seen by the observer is approximately  $(c(\omega)-v(E))\Delta t$, which arrives at the observer over a time interval equal to this distance divided by the speed of light. The cut off frequency of the synchrotron pulse is roughly the inverse of this time interval,
\begin{equation}
\omega_c \,=\, \frac{3}{4}\frac{1}{R(E)\delta (E)}\frac{1}{c(\omega_c)-v(E)}\,.
\end{equation}

But, as we have seen in the time delay discussion of the previous section, the real distance that the electron travels depends also on its energy. This leads us to modify the previous equation accordingly with the metric~\eqref{eq:dS_metric3} (with the plus sign in order to have a subluminal velocity for massive particles): the radius of the circular orbit is  
\begin{equation}
R(E)\,=\, r \,e^{k_0/\Lambda}\,.
\label{eq:radius_critical_frequency_1}
\end{equation}
Moreover, this correction changes also the distance traveled by the electron towards the observer, so we must replace $v(E) \Delta t$ by $v(E)  e^{k_0/\Lambda} \Delta t $. This modifies the equation found in the LIV case leading to the following expression
\begin{equation}
\omega_c \,=\, \frac{3}{4}\frac{1}{ r \,e^{k_0/\Lambda}\delta (E)}\frac{1}{c(\omega_c)-v(E)e^{k_0/\Lambda}}\,.
\label{eq:critical_frequency}
\end{equation}

Let us note that we have neglected  possible effects coming from the interaction in DSR, i.e. due to the deformed conservation law for the momenta. However, this contribution is completely negligible  since the modification of the result would be proportional to the product of the photon and electron momentum divided by the high-energy scale, which is insignificant due to the low energy of the emitted photon, making the previous formula the main source of modification in the synchrotron emission.

We start by obtaining the radius of the circular motion of the electrons. Since in order to consider the electromagnetic force one usually adds to the standard Casimir a term proportional to the velocity, we will consider the following action 
\begin{equation}
S\,=\,\int{\left(\dot{x}^\mu k_\mu-\frac{1}{2m} \left(C(k)-m^2\right)-e \dot{x}^\mu A_\mu(x)\right)}\,,
\label{eq:DGR_action}
\end{equation}
where the dot represents the derivative with respect the proper time $\tau$. In a future work we will show why this is the correct way to take into account the electromagnetic force in our cotangent bundle geometry formalism~\cite{Relancioa}. In the case we are considering, $A_\mu(x)$ has to represent a static magnetic field, so $A_0=0$ and $\vec{A}$ is a function of the space coordinates. Then, considering the Casimir as the squared distance in momentum space corresponding to the metric~\eqref{eq:dS_metric3} (with the plus sign) from the origin to a point $k$~\cite{Gubitosi:2013rna}
\begin{equation}
C(k)\,=\,d^2(0,k)\,=\,\Lambda^2 \arccosh \left(\cosh\left(\frac{k_0}{\Lambda}\right)-\frac{e^{-k_0/\Lambda}\vec{k}^2}{2 \Lambda^2}\right)\,,
\end{equation}
we obtain the equations of motion at first order in the power series expansion on the high energy scale
\begin{equation}
\begin{split}
\dot{x}^0\,&=\,\frac{k_0}{m}\,,\qquad \dot{\vec{x}}\,=\,-\frac{\vec{k}}{m}\left(1-\frac{k_0}{\Lambda}\right)\,,\\
\dot{k}^0\,&=\,0\,,\qquad \dot{\vec{k}}\,=\,-e\, \dot{\vec{x}}\wedge \vec{B}\,.
\end{split}
\end{equation}
From this it is easy to obtain
\begin{equation}
\ddot{\vec{x}}\,=\,\frac{e}{m} \dot{\vec{x}}\wedge \vec{B} \left(1-\frac{k_0}{\Lambda}\right)\,.
\label{eq:motion_tau}
\end{equation}

The velocity can be easily obtained from the quotient 
\begin{equation}
v\,=\, \frac{\dot{x}}{\dot{x}^0}\,=1-\frac{m^2}{2 k_0^2}-\frac{k_0}{\Lambda} \,,
\label{eq:velocity_critical_frequency}
\end{equation}
where $\dot{x}=|\dot{\vec{x}}|$ and  $v=|\vec{v}|$. From the previous expression one can check that the choice of the positive sign of the metric~\eqref{eq:dS_metric3} is the one leading to subluminal velocities.  

In order to write Eq.~\eqref{eq:motion_tau} as a function of time, we have to find the $\gamma$ factor, i.e.  the relation between the proper time and the time from the line element of the metric~\eqref{eq:dS_metric3}
\begin{equation}
d\tau\,=\,\sqrt{dt^2 - d\vec{x}^2 e^{2k_0/\Lambda}}\,=\,dt \sqrt{1- \vec{v}^2 e^{2k_0/\Lambda}}\implies \gamma \,=\,\frac{dt}{d \tau}\,\approx\,\frac{1}{\sqrt{1- \vec{v}^2 \left(1+2 k_0/\Lambda\right)}}\,\approx\,\frac{k_0}{m}\left(1-\frac{k_0}{2 \Lambda}\right)\,. 
\end{equation}
One can note that this differs from the $\gamma$ factor obtained in~\cite{Jacobson:2005bg} for a LIV scenario.

Therefore, the left side of  Eq.~\eqref{eq:motion_tau} becomes
\begin{equation}
\ddot{\vec{x}}\,=\,\frac{d}{dt}\left(\gamma \vec{v} \right)\,=\,\gamma \vec{a}+\gamma^3\left(\vec{v}\cdot\vec{a}\right) \vec{v} \,=\,\gamma \vec{a}\,,
\end{equation}
where $\vec{a}=d \vec{v}/dt$ and we have used that the zero component of the momentum is constant and that the acceleration is perpendicular to the velocity. Now we can rewrite Eq.~\eqref{eq:motion_tau} as a function of time 
\begin{equation}
\vec{a}\,=\,\frac{e}{m \gamma} \vec{v}\wedge \vec{B}  \left(1-\frac{k_0}{\Lambda}\right)\,, 
\label{eq:motion_t}
\end{equation}
so the radius of the circular motion is
\begin{equation}
a\,=\,\frac{v^2}{r}\implies r\,=\, \frac{m v \gamma}{e B} \left(1+\frac{k_0}{\Lambda}\right) \,.
\label{eq:radius_critical_frequency}
\end{equation}
Note that this is the same expression appearing in SR times a momentum dependent factor and where the $\gamma$ factor depends on the energy.

The last ingredient we need  in order to compute the critical frequency is the angle which can be still taken as in SR to be proportional to the inverse of the $\gamma$ factor, that is
\begin{equation}
\delta (E)\,=\,\frac{1}{\gamma}\,.
\label{eq:delta}
\end{equation}

Now we can substitute Eqs.~\eqref{eq:velocity_critical_frequency},\eqref{eq:radius_critical_frequency},\eqref{eq:delta} into Eq.~\eqref{eq:critical_frequency}, finding the lowest order modification to be
\begin{equation}
\omega_c \,=\,\frac{3 B e k_0^2}{2 m^3}\left(1-\frac{2 k_0}{\Lambda}\right)\,.
\label{eq:critical_frequency_subluminal}
\end{equation}

If we derive the critical frequency with respect to the energy we find that it has a maximum value 
\begin{equation}
\omega_c^{\text{max}} \,=\,\frac{ B e \Lambda^2}{18 m^3}\,,
\end{equation}
for an energy 
\begin{equation}
k_0 \,=\,\frac{\Lambda}{3}\,.
\end{equation}
Then, we can use the Crab nebula data~\cite{Jacobson:2005bg} in order to find a constraint on the high energy scale. The radiation coming from there is of $100$ MeV and the magnetic field is no larger than $0.6$ mG in the emitting region. From our result, one can check that the constrain on the high-energy scale is $\Lambda > 7.8 \times 10^{3}$ TeV. This result completely differs  with the one obtained in~\cite{Jacobson:2005bg} for the LIV case, in which the high-energy scale should be $\Lambda\gtrsim 10^{7} M_{P}$ for a deviation of $\mathcal{O}(E/M_P)$. Note that this constraint only appears when considering that high-energy particles acquire always a subluminal speed. For the superluminal case, the critical frequency can be obtained from Eq.~\eqref{eq:critical_frequency_subluminal} just changing the sign preceding $\Lambda$
\begin{equation}
\omega_c \,=\,\frac{3 B e k_0^2}{2 m^3}\left(1+\frac{2 k_0}{\Lambda}\right)\,,
\label{eq:critical_frequency_superluminal}
\end{equation}
which has not a maximum finite value, impeding us to cast a constrain.

\section{Conclusions and future prospects}
\label{sec:conclusions}
In a previous paper~\cite{Relancio:2020zok}, it was shown how to generalize a generic pseudo-Riemannian spacetime so as to include a momentum dependence on the metric, leading to a geometry in the cotangent bundle. This metric for the whole phase space allowed us to compute phenomenological aspects for the  Friedmann-Robertson-Walker universe  case when the momentum space is maximally symmetric, and in particular for de Sitter. 

In this work, we have used these results in order to seek for constraints on the scale of the deformation. We have focused on three possible evidences: a time delay for photons and a modification of the redshift and luminosity distance. For the first case, we have seen that in our proposal there is no such an effect since, although photons with different energies ``see'' different momentum dependent spacetimes and velocities, there is a cancellation of the momentum dependence in such a way that the real velocity is $1$,  independently of their energy. This result could differ if interactions play a role in the emission and detection of the photons, which is left for another work. 

Moreover, we have contrasted the redshift of active galactic nuclei obtained from x-ray and visible spectra. Since our model proposes a momentum dependent redshift, the redshift obtained for both spectra should be different. From the data, we have found that a scale of $49$ keV could be compatible with the experimental observations.

Also, we have used the luminosity distance computed from the gravitational waves source GW170817 and the associated gamma ray burst GRB170817A in order to establish a constraint on the high-energy scale. Since the energy of the gravitational waves and photons are different, following our model, there should be a difference in both quantities. In this case, we find that a deformation scale of $8$ keV is compatible with the experimental data. 

In summary, the scale parametrizing the model we have proposed is very difficult to constrain with actual data when considering this kind of cosmological phenomenology.

However, this is not the case when studying the synchrotron radiation with a space-time metric depending on momenta in such a way that electrons are always subluminal. In this case, we have seen that there is a critical frequency in the emitted  photons leading to a constrain for the de Sitter case of $\Lambda > 7.8 \times 10^{3}$ TeV when the modification starts at first order in the scale, which nonetheless it is still $\sim 10^{13}$ orders of magnitude away from the Planck scale. 

In this work we have used a particular choice of coordinates of a de Sitter momentum metric in which the momentum corrections of the Minkowski metric starts at first order in the inverse of the high-energy scale. If we had used another coordinates in which the momentum deviation 
of the flat metric started at second order (as the one considered in~\cite{Carmona:2019fwf}) we would have generally obtained different (weaker) constraints. This is related to the fact that different bases could represent different physics was considered deeply in the literature~\cite{AmelinoCamelia:2010pd}. However, the momenta we measure might change as well depending on the momentum variables, so how we should identify momenta with our measurements and if there is a physical choice of momentum variables is still an open question. In a future work we shall address this issue in detail~\cite{Relancioa}.

\section*{Acknowledgments}
This work is supported by the Spanish grants  PGC2018-095328-B-I00 (FEDER/Agencia estatal de investigación), and DGAFSE grant E21-20R. The authors would like to acknowledge the contribution of the COST Action CA18108. We acknowledge useful discussions with César Asensio, Jesús Clemente, José Manuel Carmona, José Luis Cortés, Christian Pfeifer and Licia Verde.

\end{document}